\documentclass[11pt,a4paper]{article}

\usepackage[T1]{fontenc}
\usepackage[utf8]{inputenc}
\usepackage{lmodern}
\usepackage[english]{babel}
\usepackage{microtype}
\usepackage{booktabs}
\usepackage{array}
\usepackage{tabularx}
\usepackage{geometry}
\usepackage{enumitem}
\usepackage{xcolor}
\usepackage{amsmath}
\usepackage{amssymb}
\usepackage{graphicx}
\usepackage{subcaption}
\usepackage[colorlinks=true,
            linkcolor=blue,
            urlcolor=blue,
            citecolor=blue]{hyperref}

\setlist{itemsep=2pt,parsep=0pt,topsep=4pt}
\newcolumntype{L}[1]{>{\raggedright\arraybackslash}p{#1}}
\newcommand{\brk}{\allowbreak}
\newcommand{\corpus}{https://ai-resources.eu}
\newcommand{\axis}[3]{\href{\corpus/atti-normativi/ue/#1/\#axis-#2}{#3}}

\hypersetup{
  pdftitle={Qualified Cross-References as a Verification Method:
            The Normative Environment of the EU AI Act},
  pdfauthor={Nicola Fabiano},
  pdfsubject={Legal informatics; qualified cross-reference verification},
  pdfkeywords={AI Act, EU digital acquis, legal knowledge base,
               legal informatics, cross-reference qualification,
               provenance, knowledge-base verification}
}

\title{\bfseries Qualified Cross-References as a Verification Method:\\
The Normative Environment of the EU AI Act}

\author{Nicola Fabiano\thanks{Studio Legale Fabiano, Italy. Independent Researcher on Artificial Intelligence, Data Protection, and Privacy. Expert in the EDPB's Support Pool of Experts,  Field B: Legal Expertise in New Technologies (project-based). Member, International Neural Network Society (INNS). Member, United Nations University AI Network (UNU AI Network). Member, IEEE - IEEE SA. Member, Editorial Advisory Board, Journal of Systemics, Cybernetics and Informatics (JSCI). Member, International Institute of Informatics and Systemics (IIIS). ORCID: \url{https://orcid.org/0000-0002-8188-7656}. Curator of
\url{https://ai-resources.eu}. Correspondence:
\url{https://www.fabiano.law}.}}

\date{19 August 2026}

\begin{document}
\maketitle

\begin{abstract}
Legal cross-references are commonly represented as links between instruments or provisions. For a curated legal knowledge base, the existence of a link is only the beginning of the claim: it must also state the legal character of the interaction, identify the provisions supporting it, preserve its conditions, and remain consistent when reached from either instrument.

This paper presents a provision-level model and a construction protocol for qualified cross-references, developed through a bilingual corpus of fourteen instruments surrounding Regulation (EU) 2024/1689 (the AI Act). The model distinguishes direct textual reference, bounded presumption of conformity, substantive interaction without textual reference, mediated intersection, and institutional analogy, and treats applicative interaction and definitional overlap as independent dimensions.

The methodological contribution is \emph{bidirectional inversion}: a relationship documented from act~A towards act~B is reconstructed from B's perspective against the provisions of both. Inversion is not a duplicate table but a verification operation that tests provisions, qualification, direction, and conditions before deciding how to render the relationship from either side.

Applied during construction, the protocol surfaced six incorrect article references, three inaccurate legal qualifications, and one divergence between two published descriptions of the same interaction. The corpus also shows why qualification matters: one reference to Regulation (EU) 2019/881 carries the AI Act's bounded cybersecurity presumption for high-risk systems, while related product legislation uses the same certification framework through legally distinct mechanisms. The contribution is thus a map of one regulatory environment and an explicitly specified method for making curated cross-reference knowledge bases inspectable and internally testable. A later, separately scoped audit of the converted corpus is reported among the limitations, without pooling its findings with the original measurements; limited access to the historical inputs constrains independent reproduction.
\end{abstract}


\section{Introduction}\label{sec:introduction}

An AI system embedded in industrial machinery may engage the AI Act, the
Machinery Regulation, the Cyber Resilience Act (CRA), and the European
cybersecurity certification framework. A list of citations can identify
some of those instruments. It does not, by itself, tell a reader whether a
particular connection creates a bounded presumption of conformity, a
mandatory certification route subject to preconditions, a substantive
interaction without an express reference, or only an institutional
parallel.

That difference is material. A curated cross-reference is not merely a
pointer. It is a compact legal proposition: it identifies the connected
provisions, characterizes their relationship, and may state limitations or
conditions on the resulting legal effect. Its quality therefore cannot be
established only by confirming that both target documents exist or that a
hyperlink resolves.

This paper develops that observation into a method. It reports a qualified
cross-reference map constructed within \emph{AI Resources}, a bilingual
Italian/English corpus of EU and Italian legislation relevant to the AI
Act. The map is available in interactive and tabular form at
\href{\corpus/mappa-relazioni/}{\texttt{ai-resources.eu\brk/mappa-
\brk relazioni}}, and is reproduced in Figure~\ref{fig:map}. Each
reported relationship is anchored to a fact-sheet section and to the
provisions on which its qualification rests.

The paper makes four contributions:
\begin{enumerate}[label=(\roman*)]
  \item a provision-level model in which the interaction, rather than
  either document page, is the unit of record;
  \item a separation between applicative interaction and definitional
  overlap, preventing one from being inferred from the other;
  \item bidirectional inversion as a construction and verification
  protocol for curated legal relationships; and
  \item an empirical account of applying the protocol, including the
  defects it surfaced and the structural asymmetries it exposed.
\end{enumerate}

The contribution is intentionally bounded. It does not claim to represent
the whole EU digital acquis, to automate legal interpretation, or to
estimate error rates in other corpora. It shows that qualified legal
relationships can be represented as inspectable records and that their
construction can itself function as a quality-control operation.

Section~\ref{sec:position} locates the contribution. Sections
\ref{sec:model} and~\ref{sec:protocol} define the representation and the
protocol. Section~\ref{sec:corpus} describes the study material.
Sections~\ref{sec:results} and~\ref{sec:findings} report the verification
record and substantive observations. Sections~\ref{sec:validity} and
\ref{sec:repro} address validity and reproducibility.

\section{Positioning the contribution}\label{sec:position}

Network representations of legislation establish that legal instruments
form structured systems of citations, amendments, and legal bases
\cite{koniaris2018}. Semantic edge labeling further shows that explicit
citations can be classified by purpose rather than represented as
untyped links \cite{sadeghian2018}. At the policy level, the interaction
between the AI Act and the wider EU digital framework has also been
examined as a source of regulatory complexity \cite{epinterplay2025}.
These strands establish the importance of both connectivity and
qualification.

The present contribution concerns an object that direct citation-edge
labeling does not represent. Semantic edge labeling classifies a citation that already
exists: the arc is given, and the task is to assign it a purpose. Two of
the relationship classes recorded here have no arc to label. A substantive
interaction holds where neither relevant passage names the other act, yet
the provisions operate on connected obligations; a mediated intersection
holds where the connection is carried by a third instrument invoked
independently by both sides. Neither can be produced by classifying
citations, because in neither case is there a citation between the two
acts to classify.

The difference is not only one of coverage. Extraction and labeling
answer questions about a text; a curated qualified entry makes an
assertion about a legal consequence, with conditions, and such an
assertion can be false in ways a correctly extracted citation cannot. Once
a human-authored entry says that one provision produces a particular
consequence in relation to another, the quality question is no longer
whether a citation was detected or labeled. The entry must remain
faithful to both provisions, to the direction of the interaction, and to
any conditions that bound its effect.

The EUR-Lex ELI documentation records a related limit from the
infrastructure side: its visual interface represents relationships between
whole acts, and finer resolution would require ELIs assigned at article
and clause level \cite{eurlexeli}. The provision-level model adopted here
addresses that granularity directly, while the verification lifecycle
addresses a question that arises only once assertions are authored rather
than detected.

The original object studied here is therefore the qualified interaction
record and its verification lifecycle. The map is the public rendering of
those records; bidirectional inversion is the procedure used to test them.
This framing makes the work complementary to extraction, network analysis, and regulatory interpretation rather than dependent on displacing any of
them.

\section{A model for qualified legal interactions}\label{sec:model}

\subsection{Unit of analysis}

Let a qualified interaction be represented as
\begin{equation}
  R = \langle A,B,P_A,P_B,Q,E \rangle ,
  \label{eq:record}
\end{equation}
where $A$ and $B$ are legal instruments; $P_A$ and $P_B$ are the sets of
relevant provisions in each instrument; $Q$ is the legal qualification;
and $E$ identifies the documentary evidence from which the published
renderings can be checked. The pair $\{A,B\}$ identifies the connected
instruments, not a unique legal fact: several records may concern the same
pair through different provisions or qualifications. The explanation may
be directional: the account read from $A$ need
not be a textual transposition of the account read from $B$.

At the measurement revision, a validated implementation sample represented
$E$ by the evidence locator, the language of the rendering, the corpus
revision at which it was recorded, and a content fingerprint of the row.
The sample was explicitly non-operational and covered one pair, not a
completed registry of the corpus. Fingerprints support detection of a
rendering that has been moved or edited against a recorded baseline.

Equation~\eqref{eq:record} is an abstraction of the corpus record, not a
claim that legal meaning is exhausted by a tuple. Its purpose is to make
the assertions carried by a cross-reference explicit and testable.

\subsection{Qualification}

Five categories organize most relationships in the corpus. More specific
editorial qualifications remain available within them.

\begin{table}[htbp]
\centering
\caption{Core qualification categories}
\label{tab:taxonomy}
\begin{tabularx}{\textwidth}{L{0.27\textwidth}X}
\toprule
Category & Operational criterion \\
\midrule
Direct textual reference & At least one act names the other in the
relevant normative text. \\
Bounded presumption of conformity & Satisfaction of the stated condition
supports a presumption of compliance with identified requirements, within
the bounds expressed by the provision. \\
Substantive interaction & The provisions operate on connected facts,
actors, products or obligations even though neither relevant passage names
the other act. \\
Mediated intersection & The relationship is made legible by a third act
or regime invoked by both sides; the mediator is recorded explicitly. \\
Institutional analogy & The acts exhibit parallel structures or
institutions, but the parallel is not represented as a legal reference or
legal consequence. \\
\bottomrule
\end{tabularx}
\end{table}

The categories prevent two opposite errors: calling every material
interaction a textual reference, and discarding legally relevant
interactions merely because a direct citation is absent. They also require
conditions to travel with the relationship. A presumption limited to the
requirements covered by a certificate is not recorded as an unconditional
equivalence.

\subsection{Two independent analytical axes}

For a pair of acts, the corpus evaluates two distinct questions:
\begin{align*}
  A_{ij} &= \text{documented applicative interaction between acts }i
  \text{ and }j,\\
  D_{ij} &= \text{documented overlap between concepts defined by }i
  \text{ and }j.
\end{align*}
Neither variable entails the other. Two instruments may operate together
while defining different objects, or share a definition without creating
the same operational relationship. A negative definitional assessment is
recorded as examined and not established, rather than being left
indistinguishable from an unexamined case.

\section{Bidirectional inversion protocol}\label{sec:protocol}

\subsection{Rationale}

Suppose a fact sheet for act~$A$ contains an interaction with act~$B$.
Copying the table into B's page would reproduce the assertion without
testing it. Inversion instead reconstructs the interaction from B's
perspective. Column order changes, expressions such as ``this
Regulation'' change referent, and the provision of B engaged by the
relationship must be independently identified.

The protocol turns those editorial demands into a verification sequence.

\begin{figure}[tbp]
\centering
\includegraphics[width=0.84\textwidth]{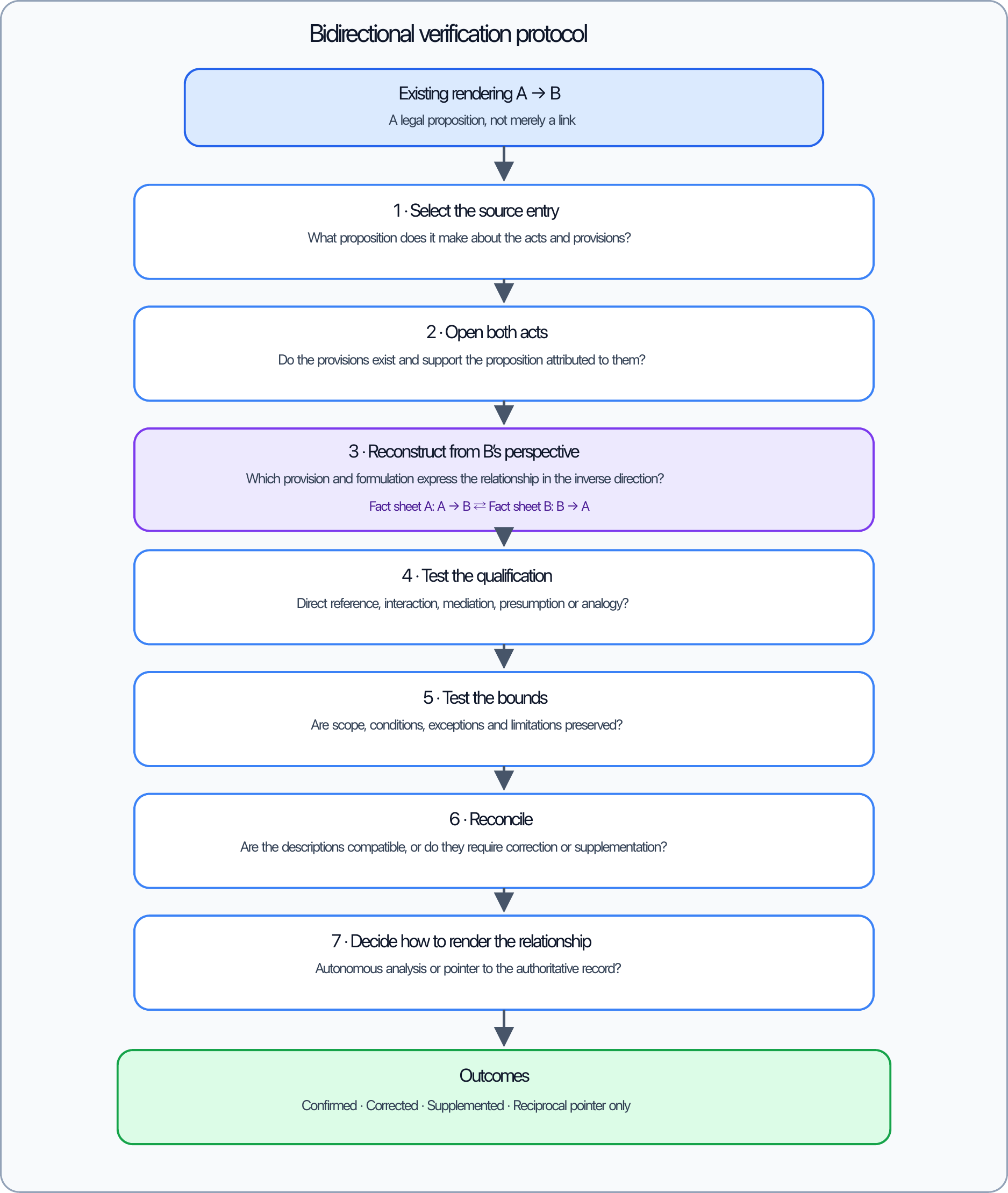}
\caption{The seven-step bidirectional verification protocol. The
relationship is reconstructed from the perspective of both acts before its
qualification, bounds, and publication form are settled. Inversion is a
verification operation, not duplication of the original rendering.}
\label{fig:protocol}
\end{figure}

Figure~\ref{fig:protocol} shows the sequence as a single decision path.
Table~\ref{tab:protocol} states the verification question attached to each
step.

\begin{table}[htbp]
\centering
\caption{Bidirectional inversion protocol}
\label{tab:protocol}
\begin{tabularx}{\textwidth}{L{0.08\textwidth}L{0.25\textwidth}X}
\toprule
Step & Operation & Verification question \\
\midrule
1 & Select the source entry & What proposition is the existing entry
making about $A$, $B$ and their provisions? \\
2 & Open both acts & Do the cited provisions exist, and do they address
the proposition attributed to them? \\
3 & Reconstruct from $B$ & Which provision and formulation express the
interaction from B's legal perspective? \\
4 & Test qualification & Is the relationship direct, substantive,
mediated, presumptive, conditional, or merely analogous? \\
5 & Test bounds & Have scope, preconditions, exceptions and limits been
preserved? \\
6 & Reconcile & Are the two descriptions compatible, or does one require
correction or additional information? \\
7 & Render & Should B contain a full autonomous analysis or a pointer to
the authoritative interaction record? \\
\bottomrule
\end{tabularx}
\end{table}

The possible outcomes are: confirmed; corrected; supplemented; or
represented only by a reciprocal pointer because the inverse perspective
adds no autonomous legal content. These are protocol outcomes, not claims
that every relationship must be legally symmetric.

\subsection{Verification and publication are separate}

Bidirectional verification does not require duplicated publication. Two
full tables describing the same legal fact can drift apart. The corpus
therefore separates \emph{reciprocity of discoverability} from duplication:
every documented interaction should be reachable from both fact sheets,
while the complete analysis is stored once unless the inverse perspective
adds distinct content.

The structural design is a non-directional registry in which the
interaction is the unit of record. The authored fact-sheet rows supply
the evidence; the registry relates their documentary renderings by position
and per-language content fingerprints. At the measurement revision this
design was represented by a validated sample, not a complete operational
registry. Provision addresses in the sample are language-independent and
distinguish enumeration, hierarchy, range and emphasis. This supports
mechanical checks of containment and focus against an index built from the
actual structure of each act; it does not establish complete corpus coverage.

\subsection{Correction lifecycle}

\begin{figure}[tbp]
\centering
\includegraphics[width=\textwidth]{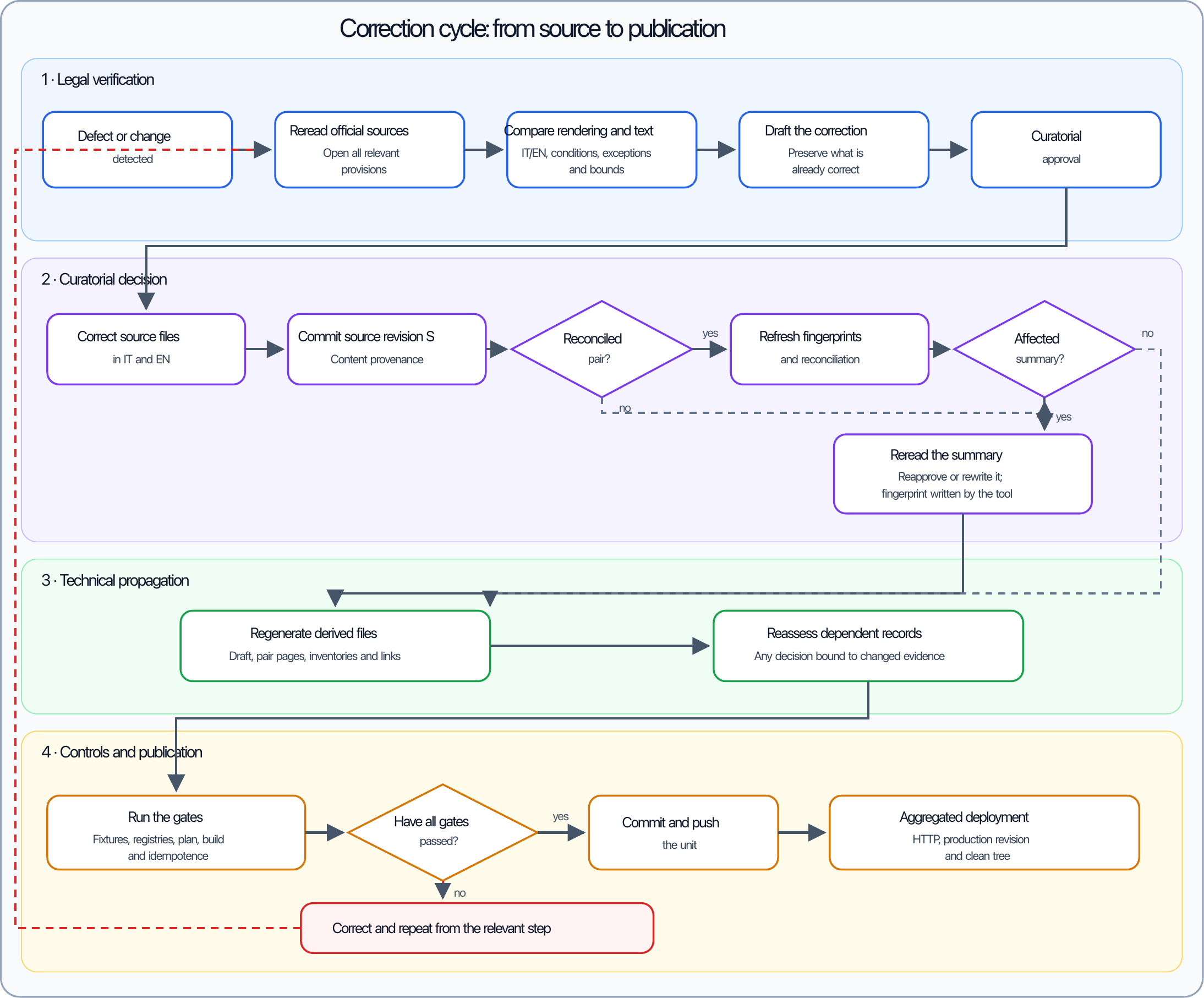}
\caption{Correction cycle from source verification to publication. Solid
arrows show the ordinary path; conditional branches identify the additional
work required for reconciled pairs and summaries; the dashed red return path
shows that a failed gate reopens the relevant step rather than being waived.}
\label{fig:correction-cycle}
\end{figure}

Verification ends with a legal assessment; it does not end the maintenance
operation. When a defect or a later source change is established, the
correction follows a controlled path from the official text to the published
derivatives. Figure~\ref{fig:correction-cycle} separates that path into legal
verification, curatorial decision, technical propagation, and controls and
publication.

The maintenance lifecycle described in this subsection was stabilized after
the measurement revision of 19 August 2026. It did not produce or
retrospectively alter the numerical results reported in this paper, which
remain anchored to that revision as stated in Section~\ref{sec:repro}.
Figure~\ref{fig:correction-cycle} describes the prescribed workflow, not
proof that every historical operation completed every gate. The separate
audit carried out after the measurement revision, and the limits of its
coverage, are reported in Section~\ref{sec:validity}.

\paragraph{Legal verification.}
The relevant official provisions are reopened and reread, including the
amending text and the base text where necessary. The existing rendering is
compared in both languages, with its conditions, exceptions and bounds. The
correction is drafted to preserve what remains accurate and is subject to
curatorial approval.

\paragraph{Curatorial decision.}
The approved wording is applied to the Italian and English source fact
sheets and committed as a source revision~$S$. If the affected pair has
already been reconciled, the evidence fingerprints and reconciliation are
refreshed against~$S$. Any curatorial summary bound to changed evidence is
reread in full and either reapproved or rewritten; its fingerprint is
written by the tool, not by hand.

\paragraph{Technical propagation.}
Derived pair pages, inventories, links and graph data are regenerated only
after the source revision. Their provenance therefore identifies the state
of the sources from which they were computed, rather than merely the later
commit that wrote them. Any dependent verification record affected by the
change is then reassessed.

\paragraph{Controls and publication.}
Fixtures, registers, structural validation, deterministic generation and the
strict site build are run as separate gates. Writing tools are subject to an idempotence gate: a second execution on
unchanged inputs must produce no change. This requirement is not, by itself,
evidence that the gate was run for each historical operation. A failed gate returns the correction to the relevant earlier step;
publication follows only after every gate has passed.

\section{Corpus and study design}\label{sec:corpus}

\subsection{Corpus}

The study uses the bilingual Italian/English legislative corpus published
by AI Resources. At the time of measurement revision, it contained fourteen
instruments: the AI Act; GDPR; Data Act; Data Governance Act; Digital
Services Act; Cyber Resilience Act; NIS2; Machinery Regulation; Product
Liability Directive; Market Surveillance Regulation; Cybersecurity Act;
Regulation (EU) 2025/37; the Digital Omnibus on AI; and Italian Law
132/2025.

Each fact sheet may contain one section for each related act, with rows
identifying provisions and qualifying their relationship. Table
\ref{tab:corpus} reports the measurements generated by the corpus tools.

\begin{table}[htbp]
\centering
\caption{Corpus measurements at the revision stated in
Section~\ref{sec:repro}}
\label{tab:corpus}
\begin{tabular}{lr}
\toprule
Measure & Value \\
\midrule
Acts & 14 \\
Cross-reference sections per language & 84 \\
Table rows per language & 315 \\
Unordered pairs with a documented relationship & 60 \\
Pairs documented from both sides & 28 \\
Pairs documented from one side only & 32 \\
\bottomrule
\end{tabular}
\end{table}

Because two cross-reference sections address multiple target acts, one with
four targets and one with two, the 84 sections encode 88 directional
declarations.

\begin{figure}[tbp]
\centering
\begin{subfigure}[t]{0.49\textwidth}
  \centering
  \includegraphics[height=6.7cm,keepaspectratio]{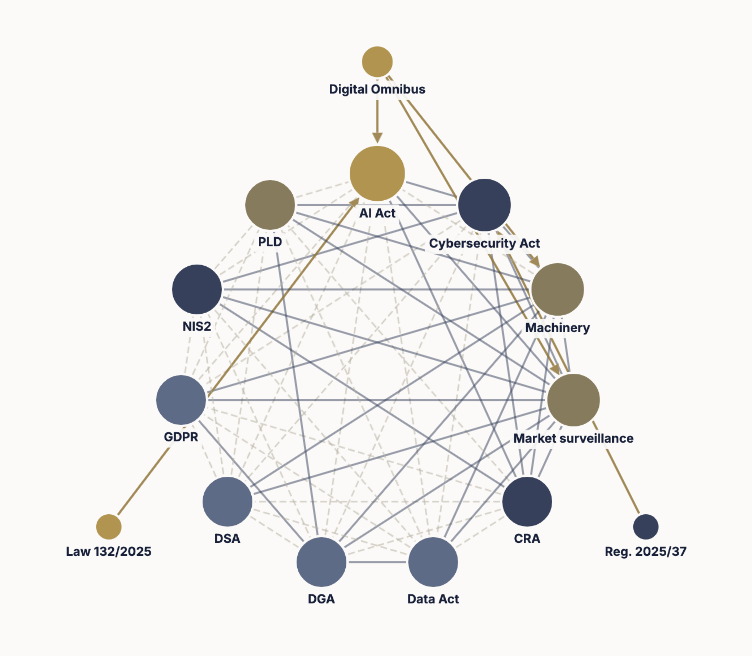}
  \caption{Concentric layout. Rings order the acts by number of
  documented relationships, placing the most connected towards the
  centre.}
  \label{fig:map-rings}
\end{subfigure}
\hfill
\begin{subfigure}[t]{0.49\textwidth}
  \centering
  \includegraphics[height=6.7cm,keepaspectratio]{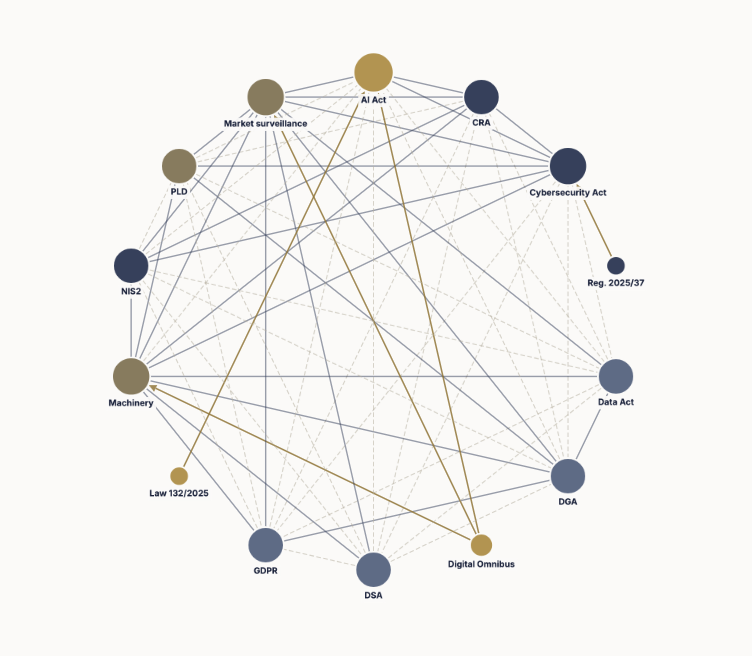}
  \caption{Circular layout. Every act occupies the perimeter, so that
  each edge can be traced individually.}
  \label{fig:map-circle}
\end{subfigure}
\caption{The qualified map, in two deterministic layouts. Each node is an
act, coloured by its prevailing domain, artificial intelligence,
cybersecurity, data and digital services, product safety, and sized by
the number of acts it is related to. Each edge is one unordered pair:
solid where the relationship is documented from both sides, dashed where
it is documented from one, and arrowed where the direction belongs to the
instrument, as with an amending act or a national law adapting to a
regulation. The rendered graph therefore carries sixty edges, one per pair
in Table~\ref{tab:corpus}, not the eighty-eight directional declarations
the underlying data contains. Both layouts are deterministic by design: a
force-directed arrangement of this graph overlaps its nodes and produces a
different figure on each load, so that two readers would not see the same
map. The interface controls and legend are cropped; screenshots taken from
the English version at the revision stated in Section~\ref{sec:repro}.}
\label{fig:map}
\end{figure}

The taxonomy is deliberately extensible. Eighty-six distinct
qualifications occur in the Italian corpus, principally as refinements of
the core categories in Table~\ref{tab:taxonomy}. The count describes the
editorial vocabulary; it is not presented as an exhaustive ontology of
legal relations.

\subsection{Questions examined}

The case study examines three questions:
\begin{description}
  \item[RQ1.] Can the legal load of a cross-reference differ materially
  from what its textual frequency suggests?
  \item[RQ2.] What defects become visible when a curated relationship is
  reconstructed from the opposite act's perspective?
  \item[RQ3.] Do applicative interaction and overlap of defined concepts
  produce the same pattern in the corpus?
\end{description}

The study is descriptive. It applies the protocol to the corpus and
reports the observations generated by that application. It does not use a
control corpus or claim a population-level error estimate.

\section{Verification results}\label{sec:results}

Twenty relationships were inverted, and twenty-two candidate additions
to definitional comparison tables were assessed. Across that work, six
incorrect article references, three inaccurate qualifications, and one
inter-sheet divergence were surfaced. Because one reference defect arose
during definitional work rather than inversion, no rate using ``six out of
twenty'' is reported.

\begin{table}[htbp]
\centering
\caption{Incorrect provision references surfaced during construction}
\label{tab:errors}
\begin{tabularx}{\textwidth}{L{0.24\textwidth}L{0.22\textwidth}X}
\toprule
Recorded reference & Corrected result & Basis of correction \\
\midrule
Art.~15 NIS2 & Art.~18 & Article~15 concerns the CSIRTs network; the
biennial report on the state of cybersecurity is in Article~18. \\
Art.~66 AI Act & Art.~67(5) & Article~66 concerns the AI Board; ENISA's
permanent membership of the advisory forum is in Article~67. \\
Art.~24 Machinery Regulation & Art.~20(9) & Article~24 concerns affixing
the CE marking; the relevant presumption is in Article~20(9). \\
Art.~3(3) Regulation 2019/1020 & Art.~11(5) & Article~3(3) defines market
surveillance; the relevant treatment of accredited certificates is in
Article~11(5). \\
Art.~30 Data Act & Art.~35(1)(d) & Article~30 concerns technical aspects
of switching; the relevant provision is in Article~35. \\
Art.~2(24) Cybersecurity Act & No corresponding definition; relevant use
at Art.~51 & Article~2 ends at point~22. ``Vulnerability'' is used in
Article~51 but is not defined at the recorded address. \\
\bottomrule
\end{tabularx}
\end{table}

The three qualification defects were different in kind. A direct textual
reference had been described as mediated; a certification obligation
dependent on a delegated act, an available scheme and an assurance level
had been stated without those conditions; and the operational objectives
listed in Article~51 of the Cybersecurity Act had been reduced to a
different security formulation. The inter-sheet divergence concerned two
descriptions of the CRA--Cybersecurity Act relationship that named
different provisions and qualifications.

These observations support a limited but consequential result: document
existence and link resolution are insufficient tests for a curated legal
cross-reference. The errors concerned the legal proposition carried by
the link, not merely its technical target.

\section{Substantive observations from the map}\label{sec:findings}

\subsection{Three parallel presumptions, and where they diverge}

Regulation (EU) 2019/881 supplies the European cybersecurity certification
framework relied on by three product instruments in the corpus. All three
attach a bounded presumption of conformity to it, and their formulations
are close enough to invite the assumption that they are equivalent. They
are not, and the qualified entries make the divergence visible.

\begin{table}[htbp]
\centering
\caption{Bounded presumptions attached to the certification framework}
\label{tab:hub}
\renewcommand{\arraystretch}{1.3}
\begin{tabularx}{\textwidth}{L{0.15\textwidth}L{0.175\textwidth}XL{0.175\textwidth}}
\toprule
Provision & Publication of the scheme's references in the OJ &
Requirements presumed satisfied & Bound of the presumption \\
\midrule
AI Act,\newline Art.~42(2) & expressly required &
cybersecurity requirements of Article~15 & insofar as covered by the
certificate or statement of conformity \\
\addlinespace
Machinery Reg.,\newline Art.~20(9) & expressly required &
Annex~III, sections 1.1.9 and 1.2.1 (protection against corruption; safety
and reliability of control systems) & insofar as covered by the
certificate or statement of conformity \\
\addlinespace
CRA,\newline Art.~27(8) & \textbf{not expressly required by
Article~27(8)} & essential cybersecurity requirements of Annex~I &
insofar as covered by the EU statement of conformity or certificate \\
\bottomrule
\end{tabularx}
\end{table}

The three provisions share a trigger (certification, or a statement of
conformity issued under a scheme adopted pursuant to Regulation (EU)
2019/881), an effect (a presumption of conformity) and a bound (only so
far as the certificate covers the requirements). They differ on one
precondition. The AI Act and the Machinery Regulation require that the
scheme's references have been published in the \emph{Official Journal};
Article~27(8) CRA does not state that condition. This is an express
textual precondition in two of the three provisions and not in the third.
The observation concerns the wording of Article~27(8) itself, not any
publication requirement that may follow from the scheme-adoption
mechanism as a whole. A representation recording all three as
``presumption of conformity, Reg.~2019/881'' would lose the distinction.

Beyond that presumption, the CRA relies on the framework in three further
distinct ways, which a single edge would collapse into one. Article~3, points~3 and~46 define
cybersecurity and cyber threat \emph{by reference} to Article~2, points~1
and~8 of Regulation (EU) 2019/881, so the CRA's own vocabulary is partly
borrowed. Article~27(9) empowers the Commission to specify by delegated
act which European schemes may be used for the purposes of the
presumption. And Articles~8(1) and 32(4) construct a different mechanism
altogether: for categories of critical products identified by delegated
act, and where an applicable scheme is available, certification at an
assurance level of at least ``substantial'' becomes a route for
demonstrating conformity, with alternative procedures where the conditions
are not met. Definitional borrowing, a bounded presumption, a delegated
specification power, and a conditional conformity route are four
qualitatively different dependencies on the same instrument.

\paragraph{Textual frequency and legal load.} The corpus permits a direct
measurement of the gap between the two. Table~\ref{tab:freq} reports
occurrences of the literal string \texttt{2019/881} across the four acts in
which it appears independently of an instrument's function as an amending
act. The remaining nine instruments contain none. Regulation (EU) 2025/37,
which amends Regulation (EU) 2019/881, names it as any amending act does:
once in the enacting terms, nine times in the recitals and once in an annex.
Those formal amendment occurrences are excluded from the comparison because
they do not express the kind of regulatory dependency examined here. The
count was produced by a script over the corpus text files, separating
enacting terms, recitals, and annexes and excluding front-matter metadata
and editorial apparatus. The Italian and English versions of the corpus
yield identical counts, so the table holds for both.

\begin{table}[htbp]
\centering
\caption{Occurrences of Regulation (EU) 2019/881 in the four
non-amending acts compared}
\label{tab:freq}
\begin{tabular}{lrrr}
\toprule
Act & Enacting terms & Recitals & Annexes \\
\midrule
NIS2 & 10 & 2 & 0 \\
CRA & 9 & 16 & 1 \\
AI Act & 1 & 2 & 0 \\
Machinery Regulation & 1 & 1 & 0 \\
\bottomrule
\end{tabular}
\end{table}

The AI Act and the Machinery Regulation each invoke the instrument once in
their enacting terms, and each derives from that single provision a
bounded presumption of conformity. NIS2 and the CRA invoke it nine or ten
times. The textual footprint differs by an order of magnitude while the
legal effect at issue is comparable in kind. A citation count ranks the AI
Act's dependency alongside incidental mentions; the qualified entry
records the consequence, its precondition and its bound. This answers RQ1
within the case studied: textual frequency alone does not determine the
legal load of a cross-reference.

The three relationships are documented in the corpus at
\axis{cra}{cra-to-cybersecurity-act}{CRA $\leftrightarrow$ Cybersecurity
Act},
\axis{macchine}{macchine-to-cybersecurity-act}{Machinery
$\leftrightarrow$ Cybersecurity Act}, and the
\href{\corpus/ai-act/\#axis-ai-act-to-cybersecurity-act}{AI Act fact
sheet}.

\subsection{A mediated accreditation intersection}

The Cybersecurity Act and the Market Surveillance Regulation do not name
one another in the text and annexes examined. Article~60(1) of the
Cybersecurity Act relies on Regulation (EC) No~765/2008 for accreditation
of conformity assessment bodies. Article~11(5) of Regulation (EU)
2019/1020 requires market surveillance authorities to take due account of
specified reports or certificates issued by bodies accredited under that
same regulation.

The map records the connection as a mediated intersection through the
accreditation regime, at
\axis{vigilanza-mercato}{vigilanza-mercato-to-cybersecurity-act}{Market
surveillance $\leftrightarrow$ Cybersecurity Act}. The qualification does
not assert that every cybersecurity certificate automatically establishes
conformity for every market-surveillance purpose; the precise effect
remains bounded by the scope and conditions of the applicable product
legislation. Naming the mediator makes both the connection and its limits
inspectable.

\subsection{Applicative and definitional structure diverge}

Twenty-two candidate definitional columns were examined. Nineteen met the
corpus criterion by identifying at least one relevant defined concept;
three did not, despite a documented applicative relationship, and each of
those three is recorded as examined with the concepts assessed and the
reason. In the Cybersecurity Act comparison table, four admitted columns
for the DSA, Data Act, DGA and PLD contained six populated cells out of
twenty-four: admission under a threshold of one relevant concept produces
sparse tables by design, and the threshold is not raised, since a single
decisive definition would then be suppressed.

The DGA and Market Surveillance Regulation provide a clear negative
assessment. The concepts examined in Article~2 DGA and Article~3 of
Regulation (EU) 2019/1020 did not establish a shared defined concept under
the corpus protocol, although the fact sheets document an applicative
interaction. This is a narrower and testable claim than saying that the
instruments share no vocabulary.

The same separation explains horizontal pairings. The CRA regulates
products with digital elements, while NIS2 imposes obligations on entities;
the regimes interact even though their defined objects differ. Sparse
definitional overlap can therefore accompany a substantial operational
relationship. This answers RQ3 for the assessed candidates without
generalizing beyond the corpus.

\subsection{Institution-specific routing of personal-data concepts}

The protocol also exposed the danger of normalizing every personal-data
reference to the GDPR. Article~41 of the Cybersecurity Act concerns ENISA,
a Union body, and invokes Regulation (EU) 2018/1725. That entry had
initially been recorded as a GDPR reference and was corrected when the
provision was read during construction. The observation illustrates why
the source and institutional context of a definition must be preserved
rather than inferred from its label.

\subsection{Structural asymmetry as corpus provenance}

Of sixty documented unordered pairs, twenty-eight were documented from
both sides and thirty-two from one side only. Fifteen one-sided cases were
accounted for by recorded design choices: seven concern the AI Act sheet,
which contains only relationships founded in its own text; four concern
directional relations such as amendment or national adaptation; and four
follow other recorded editorial decisions. The remaining seventeen
reflect publication order: later fact sheets record earlier instruments,
while earlier sheets have not invariably been updated in return.

The asymmetry is therefore not interpreted as a property of EU law. It is
provenance about the construction history of this corpus. Recording that
distinction prevents graph measures from being mistaken for unqualified
claims about regulatory importance.

\section{Validity and limitations}\label{sec:validity}

\paragraph{Corpus boundary.}
The graph represents documented interactions in one selected corpus, not
all relations within EU digital law. Missing edges are not evidence that a
legal interaction does not exist.

\paragraph{Editorial selection.}
Node degree and direction are affected by the order in which fact sheets
were added and by their inclusion rules. Network measures must therefore
be interpreted together with corpus provenance.

\paragraph{Single curator.}
The same curator created and re-examined the entries. Changing from
authoring to inversion exposed defects, but errors invisible in both modes
may remain. The study reports a construction record, not independent
inter-annotator agreement.

\paragraph{No population estimate.}
The ten surfaced defects establish that the protocol detected multiple
classes of problems in this corpus. They do not estimate the prevalence of
such problems in other knowledge bases.

\paragraph{No independent ground truth.}
Each correction is supported by the provision opened during verification,
but no external gold-standard map exists against which aggregate accuracy
can be calculated.

\paragraph{Open qualification vocabulary.}
The five categories structure the corpus, while the more specific
qualifications remain open. This preserves legally relevant detail but
does not yet provide a closed ontology suitable for automatic inference.

\paragraph{Corpus fidelity and levels of verification.}
A separate audit carried out after the measurement revision examined the
fidelity of the converted AI Act text, its HTML serialization and selected
internal link targets. It does not remeasure the inversion experiment or
revalidate the legal qualifications reported above. It illustrates why a
functioning reference, a faithful source rendering and a verified legal
assertion require distinct checks. In the reported local browser test,
parser repair masked invalid paragraph nesting; whitespace generated by the
back-reference apparatus did not demonstrate preservation of the source's
non-breaking spaces. A successful strict build does not, by itself,
establish source fidelity or legal correctness. Targeted rectifications of
typographic deviations, non-breaking-space losses, invalid paragraph
serialization and broken internal links were verified within defined local
scopes. Other defects of source fidelity and of link compatibility remain
open. The rectifications were not published in the recorded workflow, and
corpus-wide conformity was not established. Some historical comparison
inputs and original logs are no longer available in the examined locations;
the affected results are reported from retained summaries rather than
re-examined original logs. This limits both independent scrutiny of those
results and repetition of the affected checks.

These limitations define the paper's evidential scope. They do not
prevent the protocol from being replicated: another curator or corpus can
apply the same record model and inversion steps and report whether the
same classes of defect arise.

\section{Reproducibility and availability}\label{sec:repro}

The numerical results reported in this paper were produced by read-only
measurement runs over the published fact sheets. The analysis instruments
used in those runs shared one parser and comprised a reader for sections,
tables and rows; a graph census with JSON export; a qualification-coverage
census; a bilingual structural validator; an index of the acts' article
structure; a row-fingerprint utility; an anchor-proposal tool with collision
checks; and a column-insertion tool run without its writing option to enforce
the definitional threshold. Purpose-specific counts, such as the occurrences
reported in Table~\ref{tab:freq}, were produced by scripts over the same
corpus revision using the criteria stated with the result.

The measurements refer to the corpus commit
\texttt{317795c7a08fc748\brk 018538e93f3270fb\brk 9219f490}, dated
19 August 2026.

Canonical section anchors follow the pattern
\nolinkurl{#axis-{source}-to-{target}} and are stable under changes to
display headings.

The corpus and interactive map are published at
\url{https://ai-resources.eu/}. The map also provides a tabular
equivalent for keyboard navigation, assistive technology, print, and use
without JavaScript; the table is generated from the same data in the same
pass as the graph, so the design reduces the risk of independently maintained versions drifting
apart; shared generation alone is not a proof of correct output. Figure~\ref{fig:map}
reproduces the graph alone and is therefore a partial view: the
qualification of each relationship is carried by the fact-sheet entry the
edge points to, not by the edge itself. Original fact sheets, qualifications and
cross-reference content are licensed under CC~BY-SA~4.0; reproduced
normative texts remain subject to their source status.

The analysis instruments used at the measurement revision, the structural
convention sample with its negative fixtures, and the definitional coverage
register are versioned in the corpus repository at that revision. The later
maintenance utilities described in Section~\ref{sec:protocol}, including the
operations supporting the correction lifecycle, were developed after the
measurement date. That repository is not public, and no tooling archive
accompanies this preprint.

The public fact sheets allow inspection of the legal propositions currently
rendered there, but are not a frozen copy of the measurement revision.
Independent reproduction of historical counts and correction sequences
requires the corresponding source snapshot, instruments and records; a
commit identifier in a non-public repository does not itself provide that
access. The model and inversion procedure are specified here so that they
can be applied elsewhere, but methodological specification, public
inspectability and empirical reproducibility are distinct claims. The
later audit reported in Section~\ref{sec:validity} does not remove these
access limitations.

\section{Conclusion}

This paper treats a legal cross-reference as a verifiable assertion rather
than a hyperlink. The assertion has identifiable provisions, a legal
qualification, conditions, and evidence. Modeling the interaction as the
unit of record makes those elements inspectable; separating applicative
and definitional axes prevents operational proximity from being confused
with shared terminology.

Bidirectional inversion supplies the construction method. Reconstructing
an entry from the opposite act's perspective requires the curator to open
both provisions, identify direction and scope, and reconcile the two
accounts. In this corpus the process surfaced six incorrect references,
three qualification defects and one divergent duplicate description,
answering RQ2 with an observed correction record rather than a general
accuracy claim.

The substantive examples show why the additional structure matters. Three
instruments attach bounded presumptions of conformity to the same
certification framework, while their provisions differ in whether
publication of the scheme's references is expressly stated as a
precondition; the CRA depends on that framework in four qualitatively
distinct ways in total. Two acts invoke it once
in their enacting terms and derive a bounded presumption from that single
provision, while two others invoke it nine or ten times: textual frequency
and legal load diverge by an order of magnitude. And instruments that
interact operationally may share no defined concept under the stated
protocol.

The resulting scientific contribution is the combination of artifact and
method: a provision-level qualified map, an explicit provenance model, a
replicable verification protocol, and an empirical correction record.
Together they provide a basis for building curated legal knowledge bases
whose legal assertions can be traced, tested, and maintained. The subsequent
audit of the converted corpus reinforces the distinction between a functioning
reference, a faithful source rendering, and a verified legal assertion: success
at one level does not certify the others. The rectifications it prompted are
bounded local results, not a claim of full-corpus conformity or completed
publication.

\end{document}